\documentclass[%
 aip,
 amsmath,amssymb,
 reprint,%
]{revtex4-1}

\usepackage{graphicx}
\usepackage{dcolumn}
\usepackage{bm}

\usepackage[utf8]{inputenc}
\usepackage[T1]{fontenc}
\usepackage{mathptmx}
\usepackage{etoolbox}
\usepackage{subcaption}

\makeatletter
\def\@email#1#2{%
 \endgroup
 \patchcmd{\titleblock@produce}
  {\frontmatter@RRAPformat}
  {\frontmatter@RRAPformat{\produce@RRAP{*#1\href{mailto:#2}{#2}}}\frontmatter@RRAPformat}
  {}{}
}%
\makeatother
\begin{document}

\preprint{AIP/123-QED}

\title{Suppression of two stream instability in relativistic electron-ion plasmas}
\author{Vivek Shrivastav}
\author{Mani K Chettri}%
\affiliation{Department of Physics, Sikkim University, Gangtok, India, 737102}%

\author{Hemam D Singh}
\affiliation{Department of Physics, Netaji Subhas University of Technology, New Delhi-110078}%

\author{Rupak Mukherjee}
\email{rmukherjee@cus.ac.in}
\affiliation{Department of Physics, Sikkim University, Gangtok, India, 737102}%

\date{\today}

\begin{abstract}
This paper investigates the suppression of two stream instabilities in electron ion plasmas when the individual species attain relativistic velocities. This suppression of the growth rate of two stream instability is consistent even when the ions form a neutralizing background. It is found that the parameter space of the growth rate reasonably squeezes for relativistic electrons at higher plasma frequencies. We further report the suppression of the growth rate of the said instability as the ion electron mass ratio reaches the realistic limit. Our results have implications for high-energy plasmas, laser-plasma interactions, and relativistic particle beam physics, providing insights into the complex interplay of linear and nonlinear processes governing the two-stream instability. Our unified four-regime analysis extends previous understanding of how realistic mass ratios fundamentally modify relativistic suppression effects, providing essential scaling laws for high-energy plasma applications.
\end{abstract}

\maketitle

%

\section{\label{intro}Introduction}

The two-stream instability is a classical plasma phenomenon that occurs when two oppositely streaming charged particle populations interact through electrostatic perturbations \cite{krall1973principles, bittencourt2013fundamentals, davidson1975influence}. The study in two stream instability has been going on for more than a half century \cite{buneman1959dissipation}.  The earliest theoretical description of a beam-driven wave growth was provided by Bohm and Gross \cite{bohm1949theory}, showing that a beam with a narrow velocity distribution can lead to exponential amplification of plasma oscillations\cite{buneman1958instability}. As the instability develops, the growing mode can strongly interact with the plasma particles, leading to significant heating in the system\cite{koskinen2010space}. The two-stream is associated with the formation of a large, localized electric field which is parallel to the current flow or magnetic field inside plasma\cite{carlqvist1973double, block1978double, quon1976formation, iizuka1979buneman, mukherjee2019turbulence, mozer2013megavolt, alfven1986double}. It is called double layer, since it has two separated and oppositely charged layers of particles. \\

This instability plays a central role in particle energization in beam-plasma systems, laboratory plasmas\cite{takeda1991observations, zhang2023generation}, space plasmas \cite{weatherall1997modulational, alfven1939motion, bennett1934magnetically, medvedev2004long, alfven1986double, mukherjee2019study, chakraborty2023two, soosaleon2015two}, nuclear fusions \cite{thonemann1958production, bernard2023effect} and high-energy-density experiments\cite{chen1985acceleration, roth2001fast, joshi2007development}. Pierce and Heibenstreit \cite{pierce1949new} showed that two stream instability leads to the exponential amplification of electric field fluctuations and contributes to energy transfer between streams\cite{buneman1959dissipation, ishihara1980nonlinear, dawson1959nonlinear}. This energy transfer from particle streams into collective wave modes results in turbulence, anamolous transport and catastrophic breakdown of ordered behaviour in plasmas\cite{zhou2023two}.\\

Many authors have studied two stream instability extensively within the framework of cold\cite{gold1965generalized} and non-relativistic systems, \cite{cordier2000two, koide2023relativistic} typically considering only electron dynamics with a stationary ion background\cite{buneman1959dissipation, stix1992waves, krall1973principles}. While this approximation provides valuable insight into the basic instability mechanisms \cite{mukherjee2019recurrence}, it is not applicable in the case of electrons and ions moving at high speeds. In systems such as astrophysical outflows, relativistic jets and laser-driven plasmas, charged particle populations often attain speeds that are a significant fraction of the speed of light. In these regimes, relativistic effects\cite{andersson2004superfluid, samuelsson2010relativistic, bret2010multidimensional}—including Lorentz mass correction and velocity-dependent inertia—significantly modify the dispersion characteristics of the instability. Moreover, in multi-species plasmas, neglecting the motion of heavier ions can lead to inaccurate predictions, especially when beam-plasma interactions involve relativistic ions or charge-symmetric configurations.\\

The theoretical framework for relativistic two-stream instability has evolved significantly over the past decade, revealing fundamental differences from classical non-relativistic treatments \cite{samuelsson2010relativistic, bret2010multidimensional, koide2023relativistic}. Samuelsson et al. \cite{samuelsson2010relativistic} provided a comprehensive relativistic fluid analysis showing that the Lorentz factor corrections lead to substantial modifications in the dispersion relation, particularly through the $\gamma'^{-3}$ (where \(\gamma'=(1-v^2/c^2)^{-1/2}\)) scaling of the plasma response \cite{andersson2004superfluid, fried1959mechanism}. Their work demonstrated that relativistic effects not only reduce the growth rate but also shift the unstable spectral range toward lower wavenumbers, fundamentally altering the character of the instability \cite{hou2015linear, kaganovich2016band}. The cold plasma approximation in relativistic regimes has been further refined by recent studies \cite{koide2023relativistic, barkov2016relativistic}, which showed that even in extremely low-density plasmas, relativistic two-stream instabilities can occur, though with significantly modified threshold conditions compared to classical predictions \cite{bret2010exact, koide2023relativistic}. Advanced numerical techniques have complemented these analytical developments, with particle-in-cell simulations revealing complex nonlinear evolution patterns \cite{umeda2006nonlinear, brown1974two} and hybrid fluid-kinetic models providing efficient frameworks for parameter studies \cite{markidis2014fluid}. The nonlinear dynamics of relativistically intense cylindrical and spherical plasma waves have also been investigated \cite{mukherjee2018nonlinear}.\\

Contemporary research has highlighted the critical importance of multi-species effects in relativistic plasma environments \cite{hawke2013nonlinear, andersson2004superfluid, weatherall1997modulational}. The inclusion of ion dynamics becomes particularly crucial when the mass ratio $m_i/m_e$ approaches realistic values, as the ion inertia provides additional stabilization beyond relativistic corrections alone \cite{cordier2000two, davidson1975influence}. Studies of electron-positron plasmas ($m_i/m_e = 1$) have shown that symmetric mass configurations exhibit fundamentally different instability characteristics compared to electron-ion systems, with broader unstable regions and enhanced growth rates \cite{andersson2004superfluid, medvedev2004long, chakraborty2023two}. This mass-ratio dependence becomes even more pronounced in relativistic regimes, where the interplay between Lorentz corrections and species inertia creates complex parameter dependencies that cannot be captured by single-species models \cite{bret2010multidimensional, barkov2016relativistic}. Multi-stream instabilities in three-species plasmas have also been investigated \cite{bawa2009multistream, lapuerta2002multistream, weatherall1997modulational}, revealing additional complexity when multiple ion species are present, particularly in space plasma environments where H$^+$ and O$^+$ ions coexist \cite{abraham2013ion}. The role of temperature gradients and kinetic effects in electron-ion two-stream systems has been examined \cite{khalil2015kinetic}, showing that thermal corrections can significantly modify the growth rates and stability boundaries predicted by cold fluid theory.\\

The suppression mechanisms in relativistic two-stream instabilities have profound implications for high-energy astrophysical phenomena \cite{medvedev2004long, weatherall1997modulational, alfven1986double}. In relativistic jet environments,  the classical two-stream instability predictions significantly overestimate the growth rates and spectral extent \cite{medvedev2004long, bret2010multidimensional, andersson2004superfluid}. The relativistic suppression effect helps to explain the observed stability of certain astrophysical outflows that would otherwise be expected to exhibit strong turbulent behavior \cite{samuelsson2010relativistic, koide2023relativistic}. Similarly, in pulsar magnetospheres and active galactic nuclei, where relativistic electron-positron pairs are prevalent, the modified instability characteristics influence particle acceleration mechanisms and electromagnetic emission processes \cite{benavcek2024streaming, plotnikov2024kinetic, usov1987two}. The understanding of these suppression effects has become increasingly important for interpreting observational data from high-energy astrophysical sources \cite{medvedev2004long, chakraborty2023two}, particularly in the context of gamma-ray bursts where ultra-relativistic shocks generate complex plasma instability cascades \cite{brainerd2000plasma, bret2010multidimensional}. Recent theoretical work has also explored the connection between relativistic two-stream instabilities and the generation of large-scale magnetic fields in cosmological settings \cite{samuelsson2010relativistic, andersson2004superfluid}, highlighting the universal importance of these phenomena across different scales and environments\cite{comer2012cosmological}.\\

Recent advances in laser-plasma interaction experiments have provided new avenues for studying relativistic two-stream instabilities under controlled laboratory conditions \cite{zhang2023generation, chen1985acceleration, joshi2007development}. Ultra-intense laser systems can now generate relativistic electron beams with precisely controlled parameters, allowing for direct experimental investigation of the theoretical predictions \cite{huijts2022waveform, thaury2015demonstration, zhang2023generation, roth2001fast}. Such experimental capabilities are crucial for validating theoretical models and exploring the nonlinear evolution beyond the linear growth phase examined in most analytical treatments \cite{ishihara1980nonlinear, dawson1959nonlinear, zhou2023two}. The controlled laboratory environment has also enabled studies of parameter regimes that are difficult to access in astrophysical observations, including systematic investigations of mass ratio effects \cite{qin20013d, moreno2018impact} and the transition from non-relativistic to relativistic behavior \cite{hou2015linear, kaganovich2016band}. Modern laser facilities have further expanded possibilities for creating electron-positron plasmas \cite{chen1985acceleration, warwick2017experimental}, providing unique opportunities to study symmetric mass configurations under controlled conditions.\\

The computational modelling of relativistic multi-species two-stream instabilities continues to present significant challenges due to the complex interplay of relativistic corrections and multi-species dynamics \cite{imbrogno2025turbulence, dieckmann2006particle}. The choice of cold fluid theory over kinetic approaches, while limiting the scope to linear analysis, provides essential analytical insight into the fundamental scaling relationships and parameter dependencies that govern these systems \cite{krall1973principles, anderson2001tutorial, mukherjee2019coherent}. This analytical framework serves as a crucial foundation for understanding more complex kinetic and nonlinear effects that emerge in realistic plasma environments, where thermal effects and wave-particle interactions can further modify the instability characteristics \cite{davidson1975influence, koide2023relativistic, pachauri2024numerical}. Recent developments in hybrid simulation techniques have attempted to bridge the gap between fluid and kinetic descriptions \cite{barkov2016relativistic, mukherjee2018three}, incorporating relativistic corrections into multi-fluid models while capturing essential kinetic effects through closure relations \cite{lemmerz2024coupling}. The computational challenges are particularly acute when modeling realistic astrophysical systems, where the wide range of temporal and spatial scales requires sophisticated adaptive mesh refinement and load balancing strategies.\\

In this work, we systematically investigate the two-stream instability using a cold-fluid model across four distinct regimes: single species: non-relativistic and relativistic, multi-species: non-relativistic and relativistic plasmas. For each case, we analytically derive the relevant electrostatic dispersion relation and numerically compute the growth rate as a function of wavenumber and other key parameters. This unified approach highlights how relativistic corrections and ion dynamics jointly influence the onset and development of the instability.\\

This paper is divided into two parts: section \ref{non relativistic case}, where we talk about the non-relativistic case, and then the relativistic case in section \ref{relativistic case}. Both sections are further subdivided into two cases each, where we talk about single-species and multi-species cases. By single species (\ref{single species}, \ref{rel single species}), we mean that only the evolution of electrons is taken, and the ions are considered to be non-evolving. In the multispecies case (\ref{Multi Species}, \ref{rel Multi Species}), we take both electrons and ions to be evolving with time. Then we will discuss the results in section \ref{reults} and then summarize our results and conclude.

\section{\label{non relativistic case}Non-relativistic two-stream instability}

Let us consider a beam of electrons injected into a plasma. Assuming the ions in the plasma form a neutralizing background, we can neglect the ions dynamics of their own. We distinguish the electrons present in the injected beam and plasma chamber as `beam electrons' and `plasma electrons' respectively. The plasma electrons were initially at rest and had an equal number of electrons and ions to maintain quasineutrality. However, when the beam electrons start propagating through the plasma, they repel the plasma electrons outward. The beam electrons are not repelled much; because of their collective beam velocity. Excess charge moves out and it quickly resumes the charge neutrality in the space of plasma-beam interaction. Thus, we can still assume the space-charge neutrality where the initial density of the beam electrons (\(n_{0b}\)) and the plasma electrons (\(n_{0p}\)) are equal to the initial density of the ions (\(n_{0i}\)). For the injected beam to generate a wave growth, the following conditions are to be satisfied\cite{liu1994interaction}: (i) the electric field of the wave must have a component along the velocity of the beam, i.e \(\Vec{E}\cdot \Vec{v}_{0b}\ne 0\)
    where, \(\Vec{v}_{0b}\) is the beam electrons' initial velocity, (ii) the frequency of the wave (\(\omega\)) as seen by the beam electrons should be small, i.e, \(\omega\) from the frame of reference of the moving beam electrons should be nearly equal to zero. The Doppler-shifted frequancy will also be nearly zero, i.e. \(\omega-\Vec{k}\cdot\Vec{v}_{0b}\simeq 0\) where \(\Vec{k}\) is the wave vector, satisfying the Cherenkov Resonance Condition, (iii) the transfer of energy between the beam electrons and wave must happen, it can be due to the bunching of the beam electrons having different polarities with the wave.\\

To understand the bunching of beam electrons, we consider an electrostatic wave of amplitude \(A\) travelling in \(z-\)direction with a phase velocity of \(\omega/k\) as, \(\Vec{E}_z=A\cos(kz-\omega t)\Hat{z}\).
The beam electron is moving in \(z-\)direction with velocity \(\Vec{v_{0b}}=v_{0b}\Hat{z}\). In the frame of reference of the wave moving with velocity \((\omega/k)\Hat{z}\), since the Doppler-shifted frequency of the wave is zero, we get \(E_z=A\cos kz'\), where \(z'\) is the moving coordinate frame.
\begin{figure}[h!]
    \centering
    \includegraphics[width=0.8\linewidth]{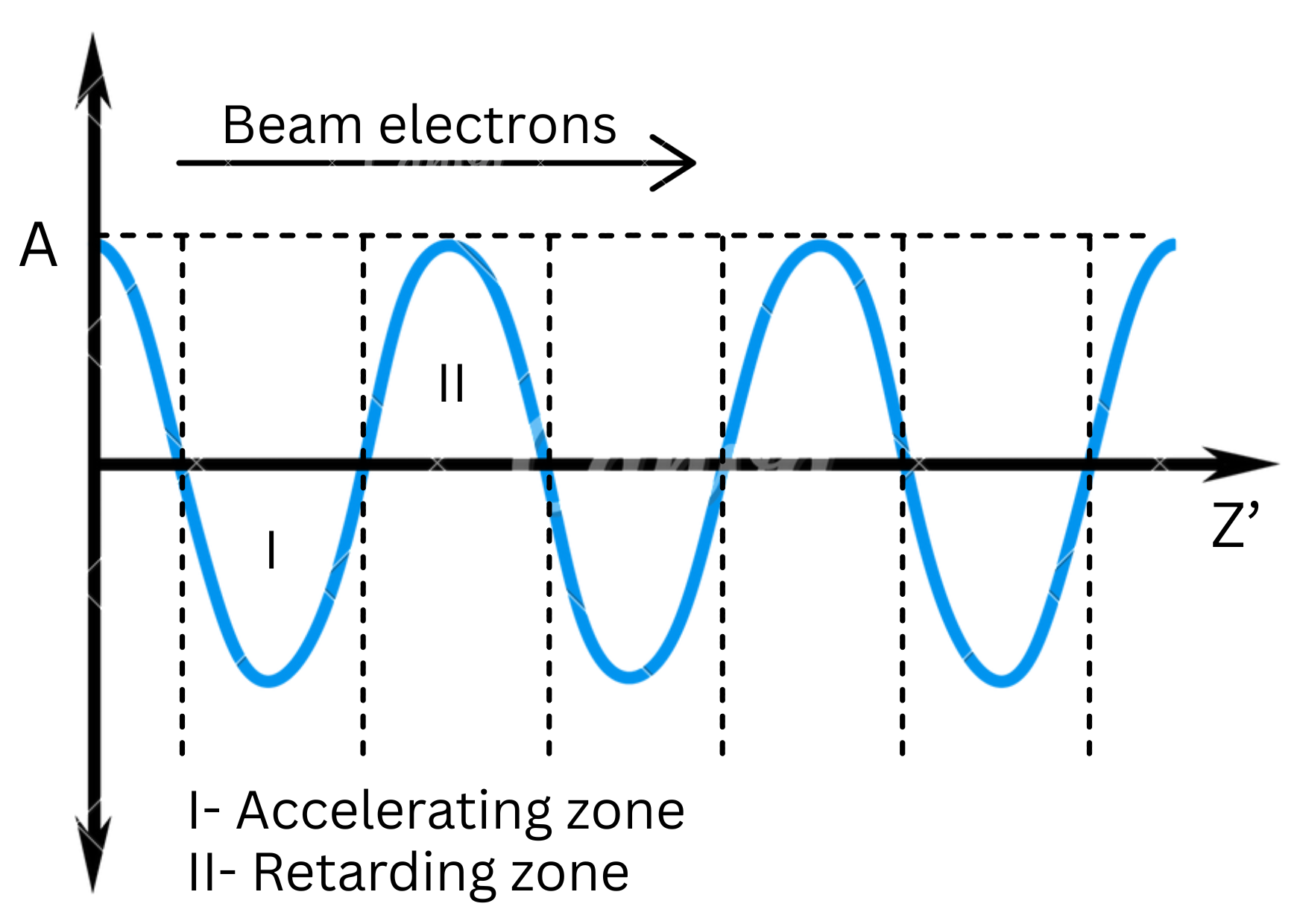} 
    \caption{Schematic diagram showing the interaction between beam electrons and wave.}
    \label{cos}
\end{figure}

The \(z\) component of the electric field of the wave is spatially positive and negative in different regions. In the negative region, the electric field has the opposite direction to the beam electrons' velocity, so it gets accelerated and is called the accelerating region. In the positive region the electric field has the same direction as the drift velocity of the beam electrons, so it gets retarded. This is called the retarding region. As depicted in the figure (\ref{cos}), we can explain the beam electron-wave interactions in three situations as follows\cite{liu1994interaction}. 
 
\begin{enumerate}
\item First case where we consider the beam electrons' velocity to be equal to the phase velocity of the wave, i.e. \(v_{0b}=\omega/k\). From the frame of the wave, the beam electrons will be seen to be stationary. In the retarding region (positive electric field), the beam electrons will feel the force backwards in \(-z\)-direction and will move backwards. In the accelerating region (negative electric field), the beam electrons will feel the force backwards in \(z\)-direction and will move in the forward direction. An equal number of electrons is transferred from the accelerating zone to the retarding zone and vice versa. Hence, no net energy is transferred from one region to another. Thus, the wave will not grow.
\item Second case when the beam electrons' velocity is less than the phase velocity of the wave, i.e., \(v_{0b}<\omega/k\), they appear to move backwards in the wave's frame. These slower electrons spend more time in the accelerating zone of the wave, where they gradually gain energy. In contrast, they pass relatively quickly through the retarding zone, experiencing less deceleration. As a result, more electrons accumulate in the accelerating zone than in the retarding zone. This leads to a net gain of energy by the electrons from the wave, causing the wave to dampen.

 \item Third case where the beam electrons' velocity is greater than the phase velocity of the wave, i.e. \(v_{0b} > \omega/k\), so they will be seen moving from the frame of reference of the wave too. Since the beam electrons have more velocity, they get accelerated more in the accelerating zone. Thus, they will quickly move to the retarding zone. However, in the retarding zone, they get retarded slowly and spend more time there. Consequently, there are fewer electrons in the accelerating zone and more in the retarding zone. Therefore, beam electrons get bunched and retarded. Hence, net energy is transferred from the beam electrons to the wave, leading to the wave growth and instability.
\end{enumerate}

\subsection{\label{single species}Single Species}

Let us consider the drift velocity of beam electrons and plasma electrons as \(v_{0b}\Hat{z}\) and \(v_{0p}(-\Hat{z})\) and the density of beam electrons and plasma electrons at equilibrium as \(n_{0b}\) and \(n_{0p}\) respectively. The electrostatic field is described as \(\Vec{E}=-\Vec{\nabla}\phi\), where the varying scalar potential \(\phi \sim e^{i(kz-\omega t)}\). The equation of motion of the beam electrons is given by
\begin{equation}
    m\left(\frac{\partial \Vec{v}_b}{\partial t}+(\Vec{v}_b.\nabla) \Vec{v}_b\right)=e\Vec{\nabla} \phi,
    \label{equation of motion of beam electrons tsi}
\end{equation}
where we have ignored the pressure and collision terms. We consider the small perturbations \(\Vec{v}_b=v_{0b}\Hat{z}+\Vec{v}_{1b}\) and apply the method of linearization. Substituting \(\frac{\partial }{\partial t} \rightarrow -i\omega\), \(\nabla \rightarrow ik\) and \(n_b=n_{0b}+n_{1b}\), equation (\ref{equation of motion of beam electrons tsi}) becomes
\begin{equation}
  -i\omega \Vec{v}_{1b}+v_{0b}ik\Vec{v}_{1b}=\frac{e}{m}ik \phi\Hat{z}.
\end{equation}
This gives
\begin{equation}
    {v}_{1bz}=-\frac{e}{m}\frac{{k}\phi}{(\omega-kv_{0b})}.
    \label{velocity of beam in tsi}
\end{equation}
When the Cherenkov resonance condition is satisfied, i.e. \(\omega-kv_{0b}\simeq0\), then \({v}_{1bz}\) becomes large. The response of electrons to the perturbation will be significant. From the linearized continuity equation, we get

\begin{equation}
   -i\omega n_{1b}+ikn_{0b}v_{1bz}+ikn_{1b}v_{0b}=0,
   \label{beam density perturbation}
\end{equation}
Substituting the value of \(v_{1bz}\) in equation (\ref{beam density perturbation}), we obtain the beam density perturbation given as
\begin{equation}
  n_{1b}=-n_{0b}\frac{e}{m}\frac{k^2\phi}{(\omega-kv_{0b})^2}.
  \label{beam electron density perturbation final in tsi}
\end{equation}
Similarly for the plasma electron density perturbation is [note: \(\Vec{v}_{0p}=-v_{0b}\Hat{z}\)]
\begin{equation}
   n_{1p}=-n_{0p}\frac{e}{m}\frac{k^2\phi}{(\omega+kv_{0b})^2}.
  \label{plasma electron density perturbation final in tsi} 
\end{equation}
Now, since we are taking ions to be non-evolving, there will be no perturbation in ion density. Therefore \(n_{1i}\simeq0\).\\
The Poisson equation is
\begin{equation}
    \nabla^2\phi=\frac{e}{\varepsilon_0}\sum_{j}n_{1j},~~\text{where j denotes beam(b) or plasma(p)}.
    \label{poisson tsi}
\end{equation}
Substituting the value of \(n_{1b}\) and \(n_{1p}\) in equation (\ref{poisson tsi}), we get

\begin{equation}
    1=\frac{n_{0b}e^2}{m\varepsilon_0}\frac{1}{(\omega-kv_{0b})^2}+\frac{n_{0p}e^2}{m\varepsilon_0}\frac{1}{(\omega+kv_{0b})^2}.
\end{equation}

If we consider the beam electron density \(n_{0b}\) and plasma electron density \(n_{0p}\) to be equal, i.e. \(n_{0b}=n_{0p}=n_{0}\), then

\begin{equation}
 1- \frac{\omega^2_{p}}{2}\left[\frac{1}{{(\omega - kv_{0b}})^2}+ \frac{1}{{(\omega + k v_{0b}})^2} \right]=0
 \label{drel}
\end{equation}
where \(\omega_p=\sqrt{\frac{2n_{0}e^2}{m\varepsilon_0}}\) is the electron plasma frequency of oscillation.
Equation (\ref{drel}) can be re-written as
\begin{equation}
    \omega^4-\omega^2(2k^2v_{0b}^2+\omega_p^2)-(\omega^2_pk^2v_{0b}^2-k^4v_{0b}^4)=0.
    \label{gamma quad single non rel}
\end{equation}
Considering \(\omega=i\gamma\), we obtain the growth rate as:
\begin{equation}
    \gamma=\sqrt{\frac{1}{2}\left[2 v^2_{0b}k^2+\omega^2_p-\sqrt{8 v^2_{0b}k^2\omega^2_p+\omega^4_p}\right]}.
    \label{quad first case}
\end{equation}
\subsection{\label{Multi Species}Multi Species}
In this section, we consider the same scenario as above except that the beam ions to be moving with the velocity \(v_{0bi}\) in \(z-\)direction and the plasma ions to be moving with the drift velocity \(v_{0pi}\) in \(-z-\)direction, thus \(v_{0bi}\Hat{z}=-v_{0pi}\Hat{z}\). 
Following the same process as that of electron dynamics in the preceding section, we obtain perturbed beam ion and plasma ion densities as
  \begin{equation}
  n_{1bi}=-n_{0bi}\frac{e}{m_i}\frac{k^2\phi}{(\omega-kv_{0bi})^2},
  \label{mul beam electron density perturbation final in tsi}
\end{equation}
and
\begin{equation}
   n_{1pi}=-n_{0pi}\frac{e}{m_i}\frac{k^2\phi}{(\omega+kv_{0bi})^2},
  \label{plasma electron density perturbation final in tsi} 
\end{equation}
respectively.
Substituting the expressions for the perturbed densities in the Poisson's equation (\ref{poisson tsi}), we can find the following relation
\begin{eqnarray}
1=&&\frac{n_{0be}e^2}{m_e\varepsilon_0}\frac{1}{(\omega-kv_{0be})^2}+\frac{n_{0pe}e^2}{m_e\varepsilon_0}\frac{1}{(\omega+kv_{0be})^2}\nonumber\\
&&+\frac{n_{0bi}e^2}{m_i\varepsilon_0}\frac{1}{(\omega-kv_{0bi})^2}+\frac{n_{0pi}e^2}{m_i\varepsilon_0}\frac{1}{(\omega+kv_{0bi})^2}\nonumber.
\end{eqnarray}

Considering \(n_{0be}=n_{0pe}=n_{0e}\) and \(n_{0bi}=n_{0pi}=n_{0i}\), we get

\begin{eqnarray}
&&1-\frac{\omega^2_{pe}}{2}\left[\frac{1}{{(\omega - kv_{0be}})^2}+ \frac{1}{{(\omega + k v_{0be}})^2} \right]\nonumber\\
&&-\frac{\omega^2_{pi}}{2}\left[\frac{1}{{(\omega - kv_{0bi}})^2}+ \frac{1}{{(\omega + k v_{0bi}})^2} \right]=0,
\end{eqnarray}

where, \(\omega_{pi}=\sqrt{\frac{2n_{0i}e^2}{m_i\varepsilon_0}}\) is the ion plasma frequency of oscillation.

\section{\label{relativistic case}Relativistic two-stream instability}
When particle velocities approach a significant fraction of the speed of light, relativistic effects fundamentally alter the dispersion characteristics of the two-stream instability. The relativistic treatment requires modifications to both the equation of motion through the relativistic momentum $\vec{P} = m\gamma'\vec{v}$ and the plasma response via the Lorentz factor $\gamma' = (1-v^2/c^2)^{-1/2}$, where we have used superscript prime to \(\gamma\) to distinguish it from growth rate. These corrections lead to the characteristic $\gamma'^{-3}$ scaling in the plasma frequency terms, which significantly suppresses the growth rate compared to classical predictions. The relativistic regime becomes essential when beam velocity $v_0 \gtrsim 0.3c$, where the Lorentz factor deviates substantially from unity and classical approximations break down.

\subsection{\label{rel single species}Single Species}
In relativistic case, wave generation can be due to Cherenkov resonance; therefore, $\omega=\vec{k}\cdot\vec{v}_{0b}\simeq\omega_p$. The equation of motion for the electron beam is given by
\begin{equation}
\frac{\partial \Vec{P}_b}{\partial t} +(\Vec{v}_b.\nabla) \Vec{P}_b = -e\Vec{E},  
\label{relativistic equation of motion}
\end{equation}
where \(\Vec{P}_b=m\Vec{v}_b\gamma'_b\) and \(\gamma'_b=1/\sqrt{1-v_b^2/c^2}\). Since \(\gamma'_b\) is a function of velocity here i.e, \(\gamma' (v_{0b}+v_{1bz})\). We will expand this \(\gamma'_b\) using Taylor series expansion as, 
\begin{equation}
      \gamma' \simeq \gamma'(v_{0b})+\frac{\partial \gamma'}{\partial v_b}\vert _{v_{0b}} v_{1bz}
    \end{equation}
        \begin{equation}
        ~~~~~~~~~~~~~~~~~~~~~~~\simeq\gamma'_{0b}+{\gamma'}_{0b}^3 \frac{v_{1bz }v_{0b}}{c^2},
    \label{gamma dash}
    \end{equation}
where \(\gamma'_{0b}=1/\sqrt{1-v_{0b}^2/c^2}\).
Linearizing the term \(\gamma'_b\Vec{v}_b\) and using equation (\ref{gamma dash}), we can write 

            \begin{equation}
        {\gamma'} \Vec{v}_b=\left[{\gamma'}_{0b}v_{0b}+{\gamma'}_{0b}^3v_{1bz}\right]\Hat{z}.
    \end{equation}
    Using the above form of \({\gamma'}\), linearized equation of motion from equation (\ref{relativistic equation of motion}) becomes
    \begin{equation}
        m\left[\frac{\partial}{\partial t}({\gamma'}_{0b}^3v_{1bz})+v_{0b}\frac{\partial}{\partial z}({\gamma'}_{0b}^3v_{1bz})\right] = eik\phi,
    \end{equation}
    which gives
    \begin{equation}
        v_{1bz}=\frac{-ek\phi}{m(\omega-kv_{0b}){\gamma'}_{0b}^3}.
        \label{vibz}
    \end{equation}
From the equation of continuity, the perturbed densities for the electron beam and plasma are given as 
 
    \begin{equation}
        n_{1b}=\frac{-n_{0b}ek^2\phi}{m{\gamma'}^3_{0b}(\omega-kv_{0b})^2},
        \label{n1b}
    \end{equation}
    and 
    \begin{equation}
        n_{1p}=\frac{-n_{0p}ek^2\phi}{m{\gamma'}^3_{0p}(\omega+kv_{0b})^2},
    \label{n1p}
    \end{equation}
   respectively. Substituting these expressions in the Possson's equation, we can find the following relation

    \begin{equation}
    1=\frac{n_{0b}e^2}{m\varepsilon_0}\frac{1}{(\omega-kv_{0b})^2{\gamma'}^3_{0b}}+\frac{n_{0p}e^2}{m\varepsilon_0}\frac{1}{(\omega-kv_{0p})^2{\gamma'}^3_{0p}}.
    \end{equation}
    Considering \(n_{0b}=n_{0p}\) and \(\Vec{v}_{0p}=-\Vec{v}_{0b}\) (thereby \({\gamma'}_{0b}={\gamma'}_{0p}\)), the above relation becomes 
        \begin{equation}
       1=\frac{\omega_p^2}{2{\gamma'}^3_{0b}}\left[\frac{1}{(\omega-kv_{0b})^2}+\frac{1}{(\omega+kv_{0b})^2}\right].
        \label{dispersion rel tsi fluid}
    \end{equation}
This can be re-written as 
\begin{equation}
    \omega^4-\omega^2\left(2k^2v^2_{0b}+\frac{\omega_p^2}{{\gamma'}_{0b}^3}\right)-\left(\frac{\omega_p^2k^2v^2_{0b}}{{\gamma'}_{0b}^3}-k^4v_{0b}^4\right)=0.
    \label{gamma quadratic relativisic single species}
\end{equation}

   
    \subsection{\label{rel Multi Species}Multi-species}
In this section, we consider the same scenario as above except that the beam ions to be moving with the velocity \(v_{0bi}\) in \(z-\)direction and the plasma ions to be moving with the drift velocity \(v_{0pi}\) in \(-z-\)direction, thus \(v_{0bi}\Hat{z}=-v_{0pi}\Hat{z}\). 
Following the same process as that of electron dynamics in the preceding section, we obtain perturbed beam ion and plasma ion densities as

    \begin{equation}
        n_{1bi}=\frac{-n_{0bi}ek^2\phi}{m{\gamma'}^3_{0bi}(\omega-kv_{0bi})^2}
        \label{n1bi}
    \end{equation}
   and
    \begin{equation}
        n_{1pi}=\frac{-n_{0pi}ek^2\phi}{m{\gamma'}^3_{0pi}(\omega+kv_{0bi})^2},
    \label{n1pi}
    \end{equation}
respectively.
    Substituting the expressions for the perturbed densities in relevant Poisson's equation, we can find the following dispersion relation

        \begin{eqnarray}
1=&&\frac{\omega_{pe}^2}{2{\gamma'}^3_{0be}}\left[\frac{1}{(\omega-kv_{0be})^2}+\frac{1}{(\omega+kv_{0be})^2}\right]\nonumber\\
&&+\frac{\omega_{pi}^2}{2{\gamma'}^3_{0bi}}\left[\frac{1}{(\omega-kv_{0bi})^2}+\frac{1}{(\omega+kv_{0bi})^2}\right].
\label{dispersion rel tsi fluid}
\end{eqnarray}

\section{\label{reults}Results}
In order to investigate the beam driven wave growth through electrostatic perturbations, we numerically solved the dispersion relations $\omega(k)$ by considering $\omega=i\gamma$ for maximum growth rate. The complex dispersion relations were solved using SciPy's root-finding algorithm, searching for solutions in the complex frequency plane by minimizing the residual of each equation. The parameter space was explored with wavenumber $k \in [0.01, 10.0]$ and beam velocities $v_0 \in [0.1, 0.95]$, with the relativistic Lorentz factor $\gamma'_0 = (1-v_0^2/c^2)^{-1/2}$ computed to maintain accuracy at high velocities. Only solutions with positive growth rates ($\gamma > 0$) corresponding to unstable modes were retained. The numerical results for single species and multi species for both non-relativistic and relativistic cases are presented in the proceeding sections.

\subsection{\label{results- single species}Single Species}

\begin{figure}[ht]
    \centering
    \begin{subfigure}[b]{0.4\textwidth}
        \includegraphics[width=\linewidth]{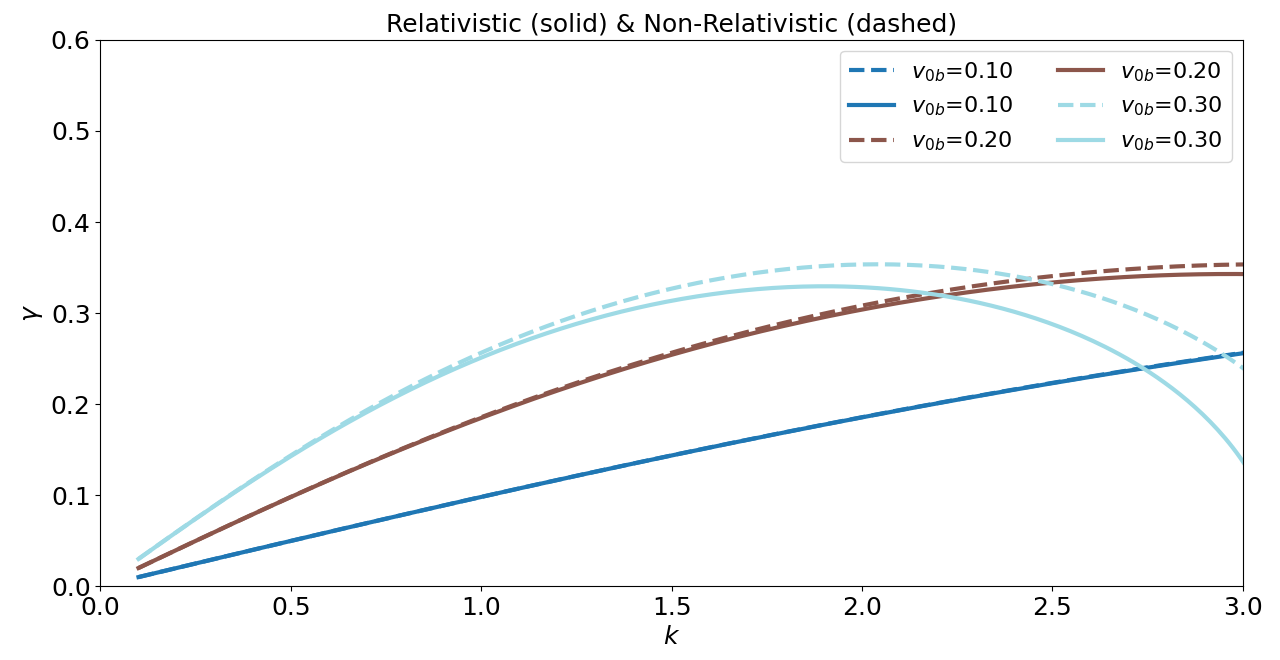}
    \end{subfigure}
    \hfill
    \begin{subfigure}[b]{0.4\textwidth}
        \includegraphics[width=\linewidth]{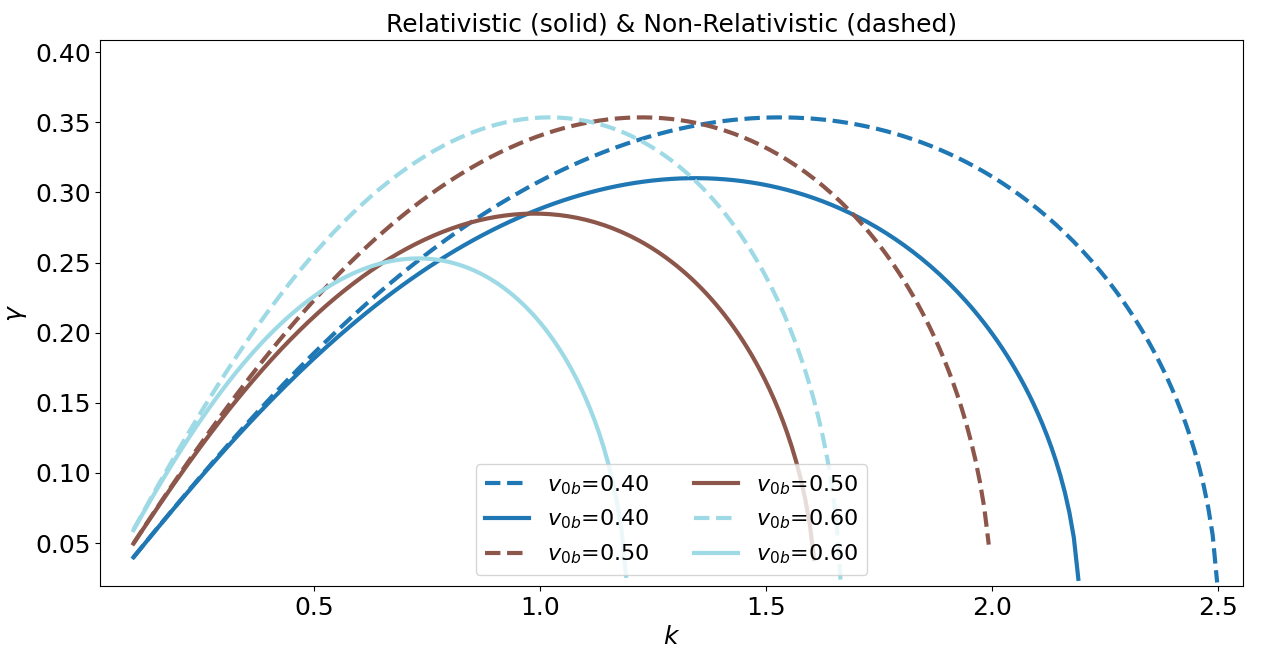}
    \end{subfigure}
     \hfill
    \begin{subfigure}[b]{0.4\textwidth}
        \includegraphics[width=\linewidth]{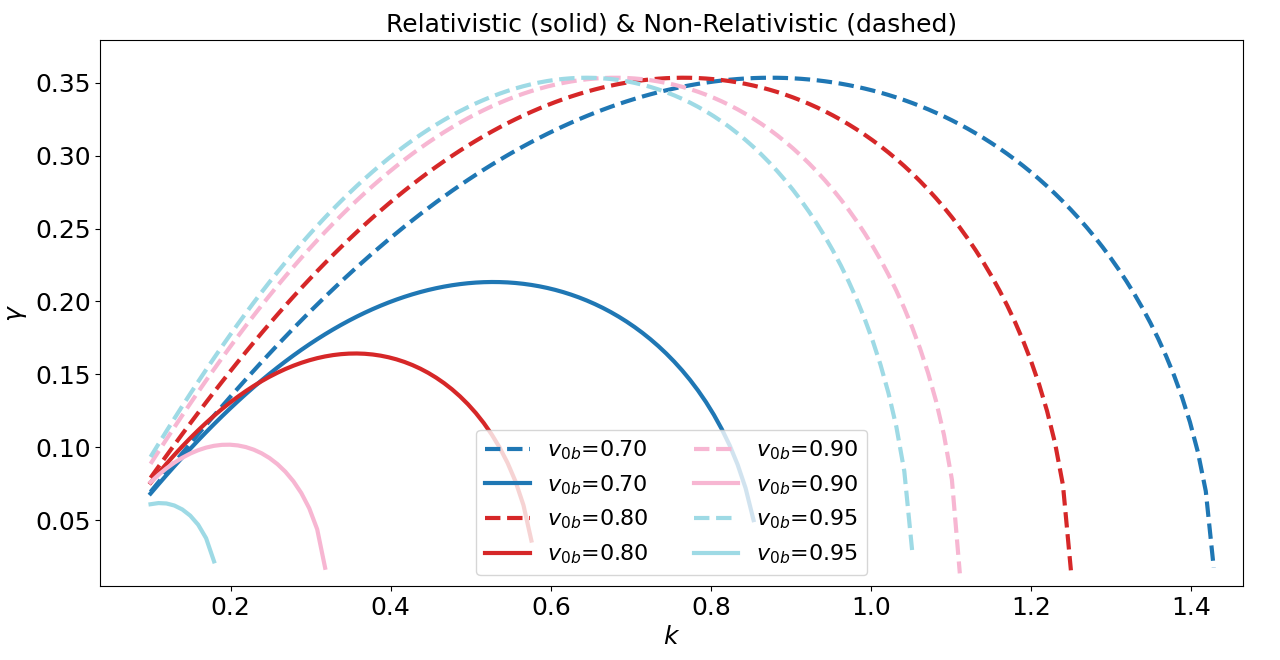}
    \end{subfigure}
 \caption{ Growth rate \(\gamma(k)\) with \(v_{0b}\in [0,0.95]\) for relativistic (solid lines) and non-relativistic cases (dashed lines). As $v_{0b}$ increases, relativistic effects become significant, reducing the growth rate and shifting the peak response.}
    \label{single species growth rate comparison}
\end{figure}

The figure (\ref{single species growth rate comparison}) presents the growth rate $\gamma(k)$ represented by equations (\ref{gamma quad single non rel}) and (\ref{gamma quadratic relativisic single species})  as a function of the wavenumber $k$ for a single-species plasma, comparing relativistic (solid lines) and non-relativistic (dashed lines) cases across various beam velocities $v_{0b} \in [0.1,0.95]$, where it is normalized by the speed of light c. At low velocities (e.g., $v_{0b} = 0.1$), both relativistic and non-relativistic curves nearly overlap, confirming that relativistic corrections are negligible in this regime. As $v_{0b}$ increases, the relativistic growth rates diverge from their non-relativistic counterparts, showing an overall reduced growth, with maximum growth happening at low values of \(k\) as reported by Samuelsson et al\cite{samuelsson2010relativistic} and Dieckmann et al\cite{dieckmann2006particle}. This suppression effect becomes particularly pronounced at higher velocities (e.g., $v_{0b} \geq 0.4$), where the relativistic curves exhibit significantly reduced peak growth rates and narrower instability bands in $k$. Meanwhile, the non-relativistic case maintains a wider and stronger instability window, even unrealistically so at $v_{0b} \rightarrow 1$, which highlights the inaccuracy of ignoring relativistic inertia in high-speed plasma beam studies.\\

\begin{figure}[h]
    \centering
    \begin{subfigure}[b]{0.4\textwidth}
        \includegraphics[width=\linewidth]{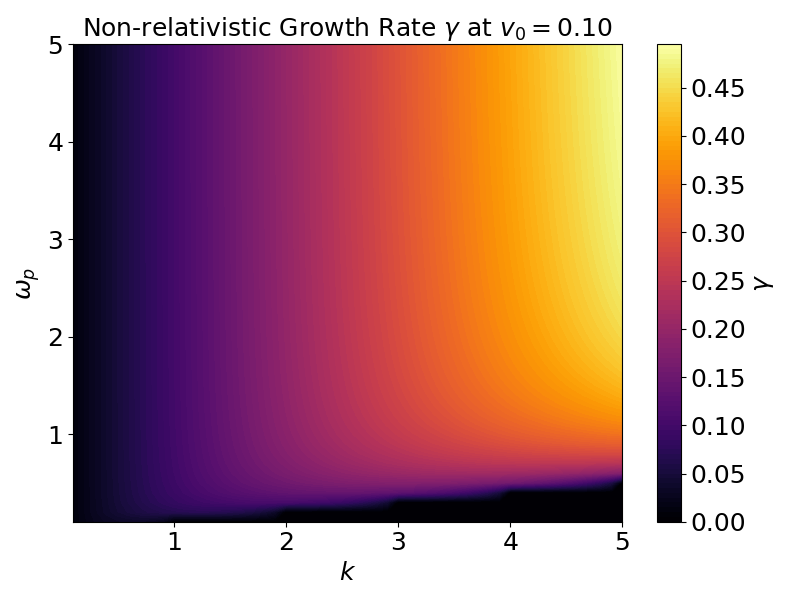}
        \label{4a}
    \end{subfigure}
    \hfill
    \begin{subfigure}[b]{0.4\textwidth}
        \includegraphics[width=\linewidth]{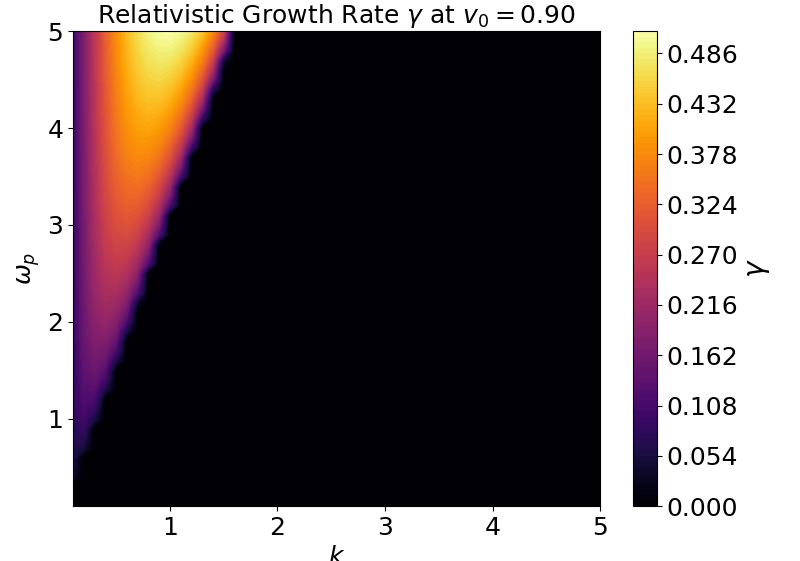}
        \label{4b}
    \end{subfigure}
    \caption{Growth rate $\gamma$ vs \((k,\omega_p)\) for a fixed drift velocity \(v_0=0.1\), non-relativistic (top panel) and \(v_0=0.9\), relativistic (bottom panel). Brighter regions indicate stronger wave growth. Relativistic inertia suppresses the instability compared to the non-relativistic case.}
    \label{compare}
\end{figure}

The two color plots shown in figures (\ref{compare}a) and (\ref{compare}b) illustrate the growth rate $\gamma$ as a function of the wavenumber $k$ and plasma frequency $\omega_p$, under non-relativistic ($v_{0b} = 0.10$) and relativistic ($v_{0b} = 0.90$) conditions, respectively. In the non-relativistic case (figure \ref{compare}a), the instability spans a broad region of parameter space \((k,\omega_p)\), with growth rates increasing steadily as both $k$ and $\omega_p$ rise. This widespread and continuous instability profile confirms that even low-velocity beams can efficiently drive two-stream growth across a wide spectral range. In contrast, the relativistic case (figure \ref{compare}b) reveals that instability is highly localized: growth occurs only for small $k$ and moderately large $\omega_p$, with a sharp cutoff at higher wavenumbers. This suppression aligns with the expected relativistic inertial effects, where the increased effective mass of the beam particles (via the Lorentz factor $\gamma'_{0b}$) reduces their responsiveness to perturbations. These observations are fully consistent with the trends observed in figure (\ref{single species growth rate comparison}), which plots $\gamma(k)$ for multiple values of $v_{0b}$: as $v_{0b}$ increases, the relativistic growth curves are both lower in amplitude and narrower in $k$-range compared to their non-relativistic counterparts. The comparison across both figures clearly demonstrates how relativistic dynamics suppress the instability window, particularly at large $k$, reinforcing the critical role of Lorentz corrections in high-speed plasma beam systems.
\begin{figure}[h!]
    \centering
    \includegraphics[width=0.4\textwidth]{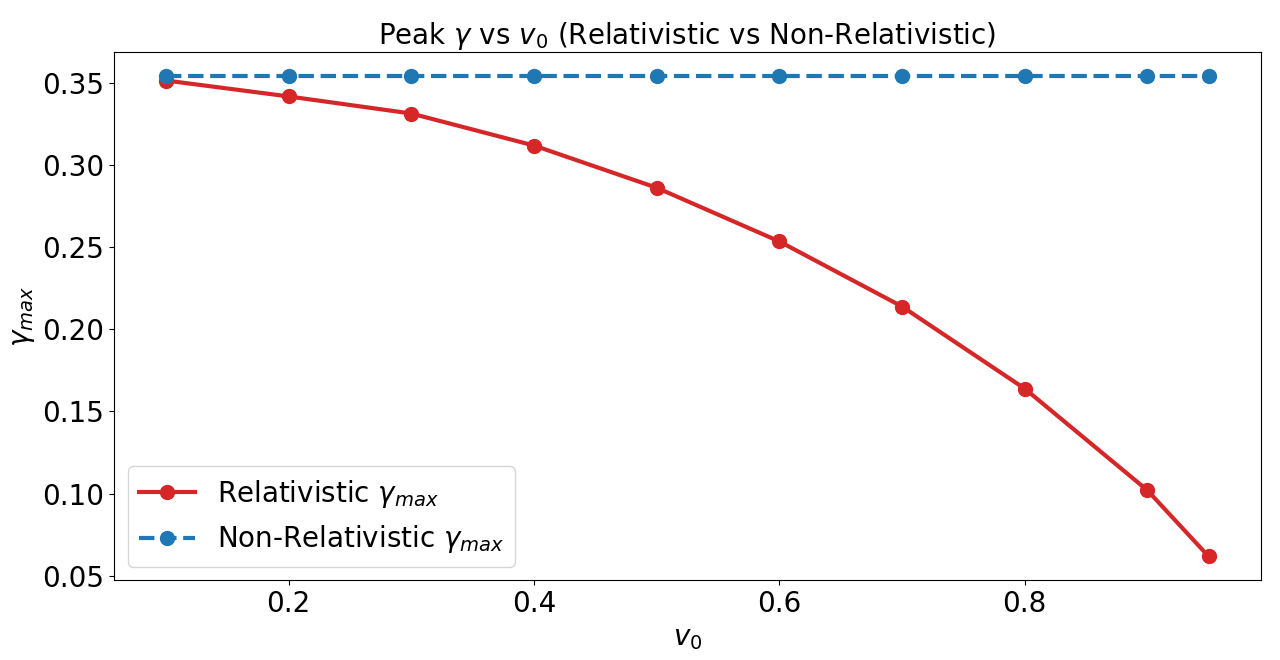}
    \caption{Maximum growth rate \( \gamma_{\text{max}} \) as a function of normalized beam velocity \( v_0 \) for a single-species cold plasma experiencing two-stream instability.}
    \label{gammavsvelocity}
\end{figure}
\\

In order to understand the relation between the beam velocity and the wave growth rate, figure(\ref{gammavsvelocity}) presents the maximum growth rate $\gamma_{\mathrm{max}}$ of the two-stream instability as a function of beam velocity $v_{0b}$ for a cold, single-species plasma. The solid red curve includes relativistic effects through the Lorentz factor \( \gamma_0^3 \), and shows that the growth rate decreases noticeably at higher beam velocities. This drop reflects the increasing inertia of the beam, which stabilizes the system. On the other hand, the dashed blue curve shows the non-relativistic prediction, where the growth rate stays nearly constant and fails to capture this stabilizing behavior. The comparison clearly shows the importance of including relativistic corrections when dealing with high-speed beams. These findings are consistent with classical results in beam-plasma systems \cite{bret2010exact}, where relativistic corrections are known to significantly alter instability thresholds and growth rates \cite{krall1973principles, fried1959mechanism}.

\begin{figure}[h!]
    \centering
    \begin{subfigure}[b]{0.4\textwidth}
        \includegraphics[width=\linewidth]{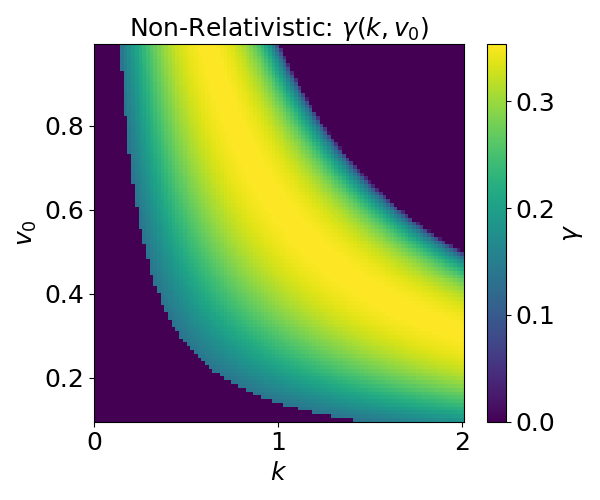}
        \label{5a}
    \end{subfigure}
    \hfill
    \begin{subfigure}[b]{0.4\textwidth}
        \includegraphics[width=\linewidth]{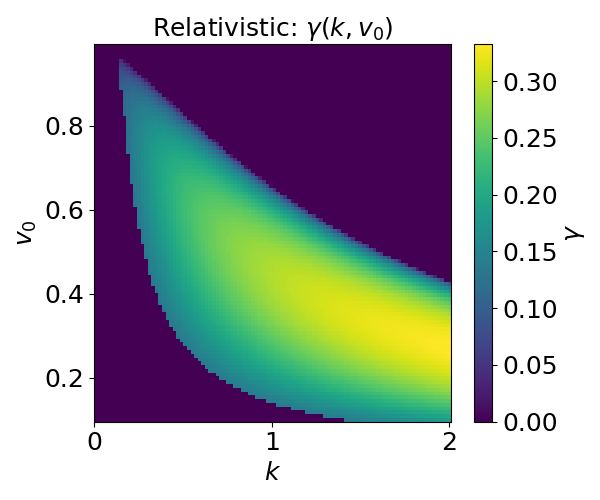}
            \label{5b}
    \end{subfigure}
    \caption{Growth rate $\gamma$ versus \((k, v_{0})\) for a. non-relativistic (top panel) and b. relativistic (bottom panel). Instability is suppressed at high $v_0$ and large $k$, narrowing the unstable region. Relativistic inertia reduces both maximum growth and spectral width.}
    \label{v0-k-contour}
\end{figure}

Figure (\ref{v0-k-contour}) displays the two-dimensional distribution of the growth rate $\gamma$ as a function of wavenumber $k$ and beam velocity $v_0$ for a single-species cold plasma system, comparing non-relativistic and relativistic regimes. 
In the non-relativistic case, the instability is observed over a broad range of both $k$ and $v_0$. As the beam velocity increases, the growth rate strengthens across the spectrum, and the unstable region expands toward higher wavenumbers. This behavior is consistent with classical theory, which predicts that faster beams resonate more strongly with plasma waves, thereby driving more vigorous and widespread instability. In contrast, the relativistic case reveals a starkly different picture. While instability is still present at moderate beam velocities, the growth rate becomes increasingly confined to lower $k$ values as $v_0 \to 1$. This spectral narrowing is a direct consequence of relativistic effects: the increased effective mass of the streaming particles (via the Lorentz factor $\gamma_0$) reduces their susceptibility to electric field perturbations. As a result, the plasma becomes more stable at high velocities, and the instability is both weakened and localized in $k$-space. This result is consistent with Hou et al\cite{hou2015linear} and Kaganovich et al \cite{kaganovich2016band}.

\subsection{\label{results- multi species}Multi Species}

\begin{figure}[h!]
    \centering
    \begin{subfigure}[b]{0.4\textwidth}
        \includegraphics[width=\linewidth]{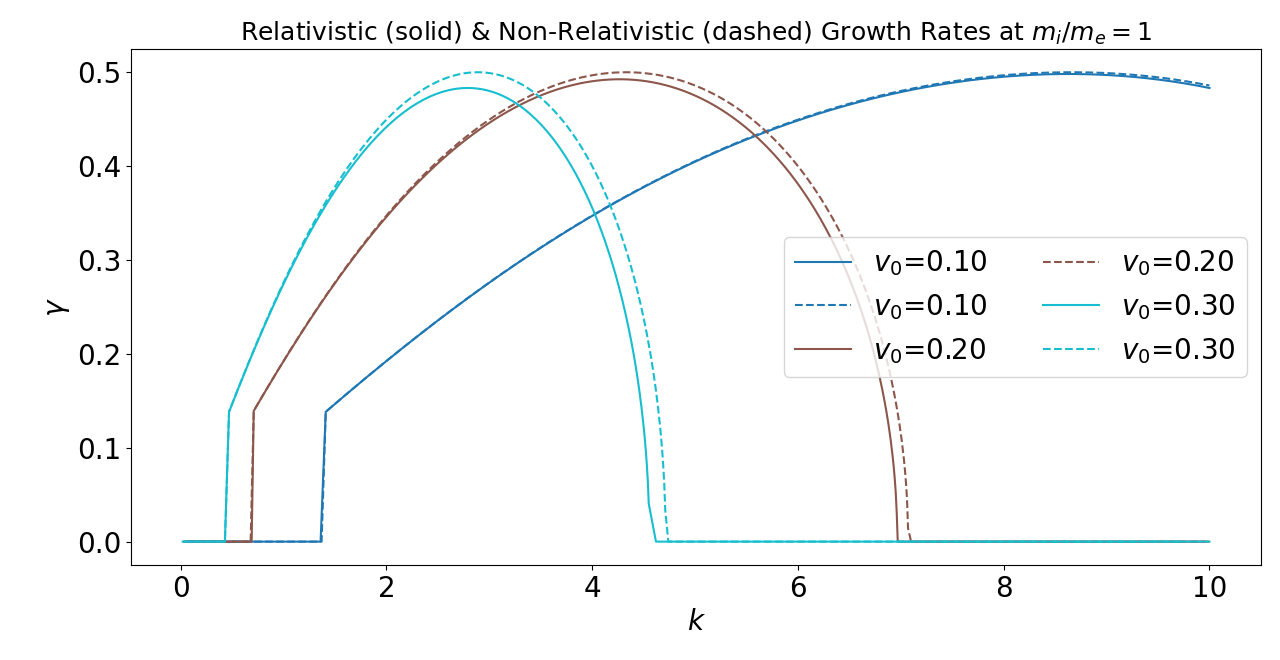}
    \end{subfigure}
    \hfill
    \begin{subfigure}[b]{0.4\textwidth}
        \includegraphics[width=\linewidth]{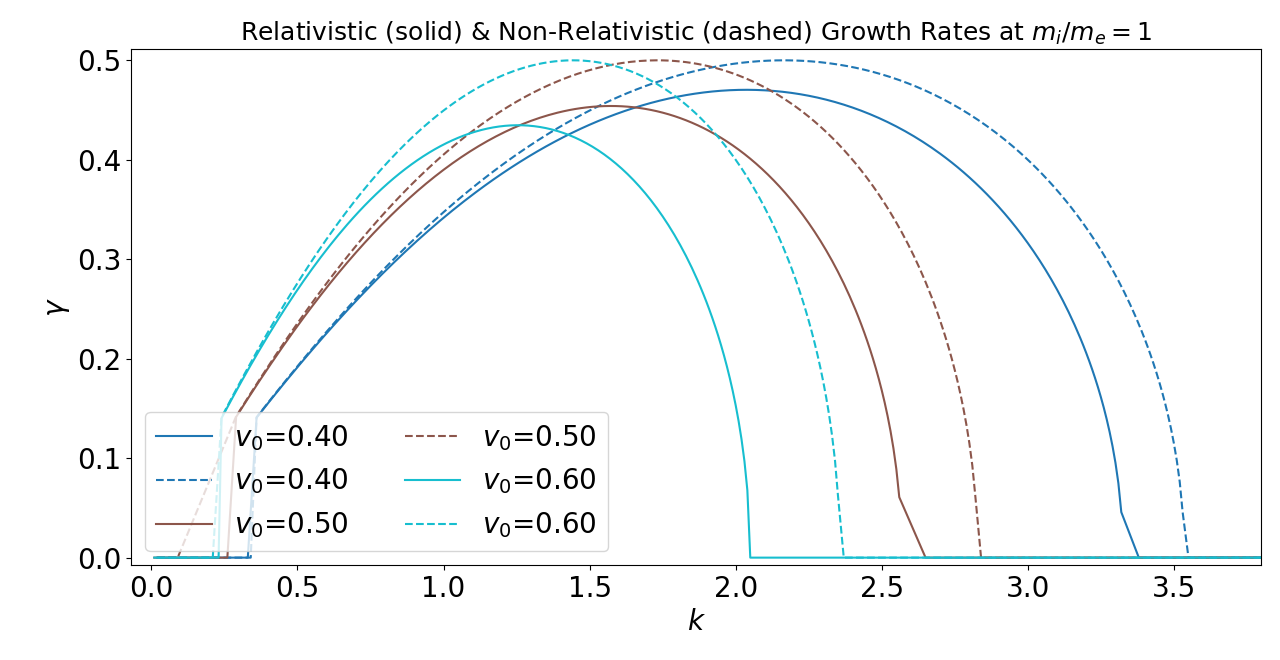}
    \end{subfigure}
     \hfill
    \begin{subfigure}[b]{0.4\textwidth}
        \includegraphics[width=\linewidth]{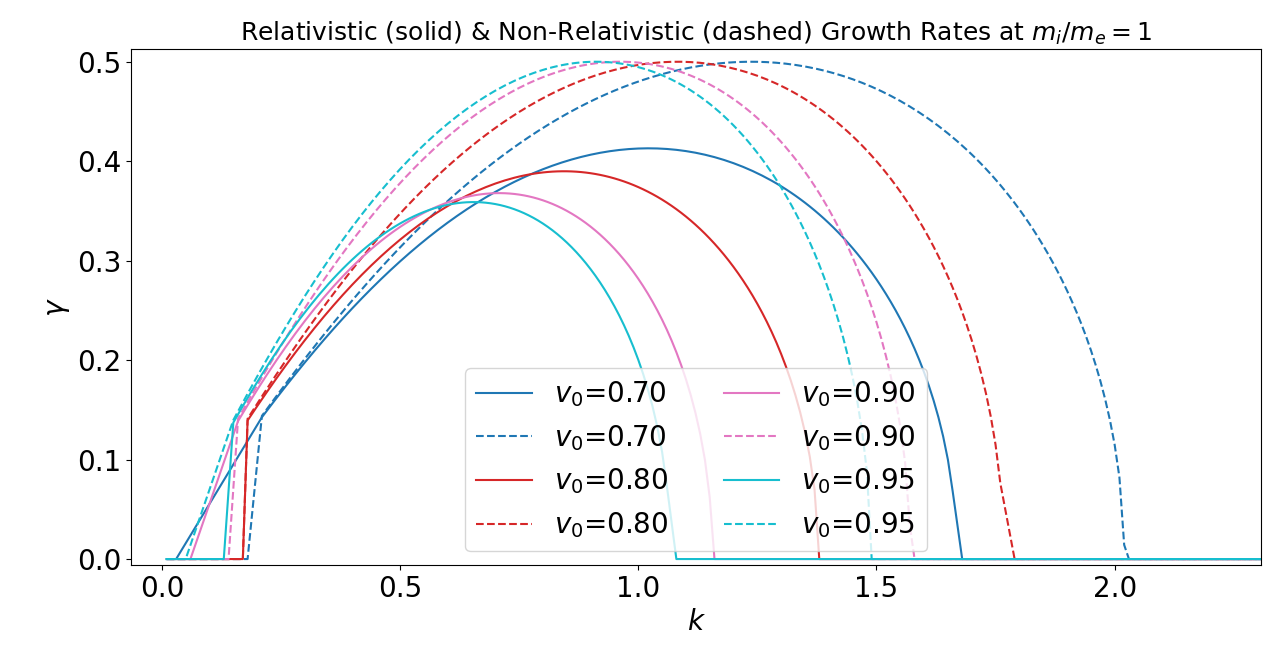}
    \end{subfigure}
 \caption{Growth rate $\gamma$ as a function of wavenumber $k$ for symmetric $m_i/m_e = 1$. Relativistic (solid lines) and non-relativistic (dashed lines) results are shown for a range of drift velocities $v_0 \in [0.1, 1.0]$.}
    \label{gamma_mi_me_1}
\end{figure}
\begin{figure}[h!]
    \centering
    \begin{subfigure}[b]{0.4\textwidth}
        \includegraphics[width=\linewidth]{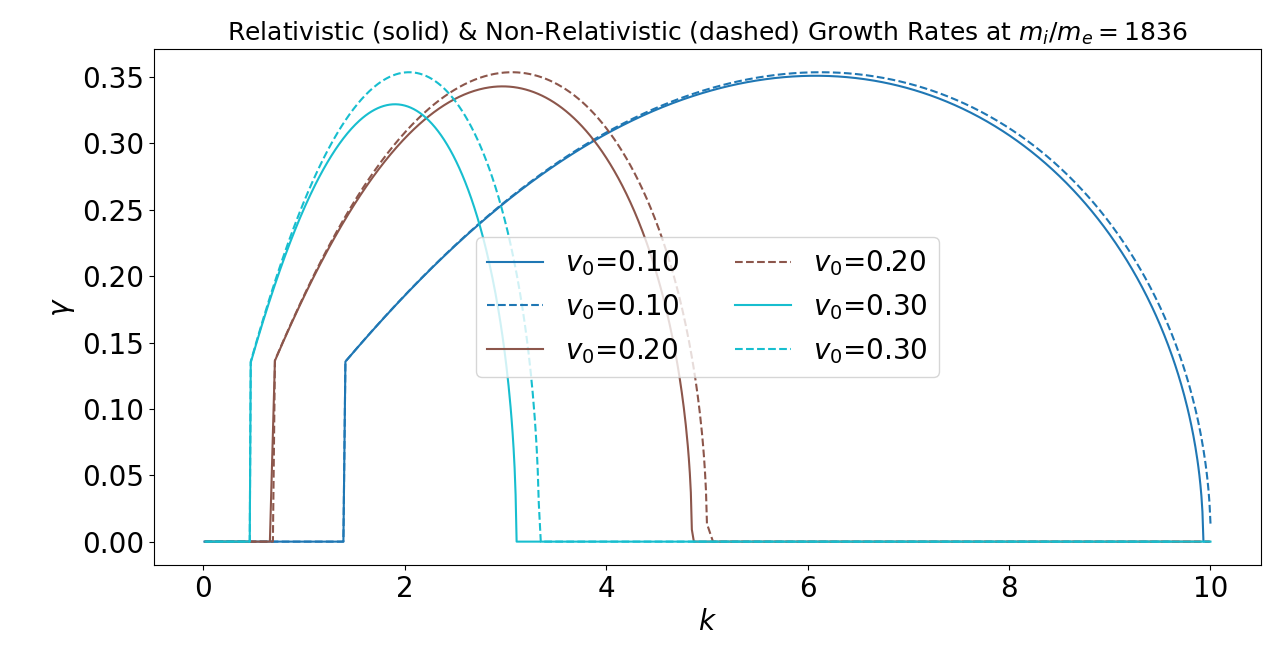}
    \end{subfigure}
    \hfill
    \begin{subfigure}[b]{0.4\textwidth}
        \includegraphics[width=\linewidth]{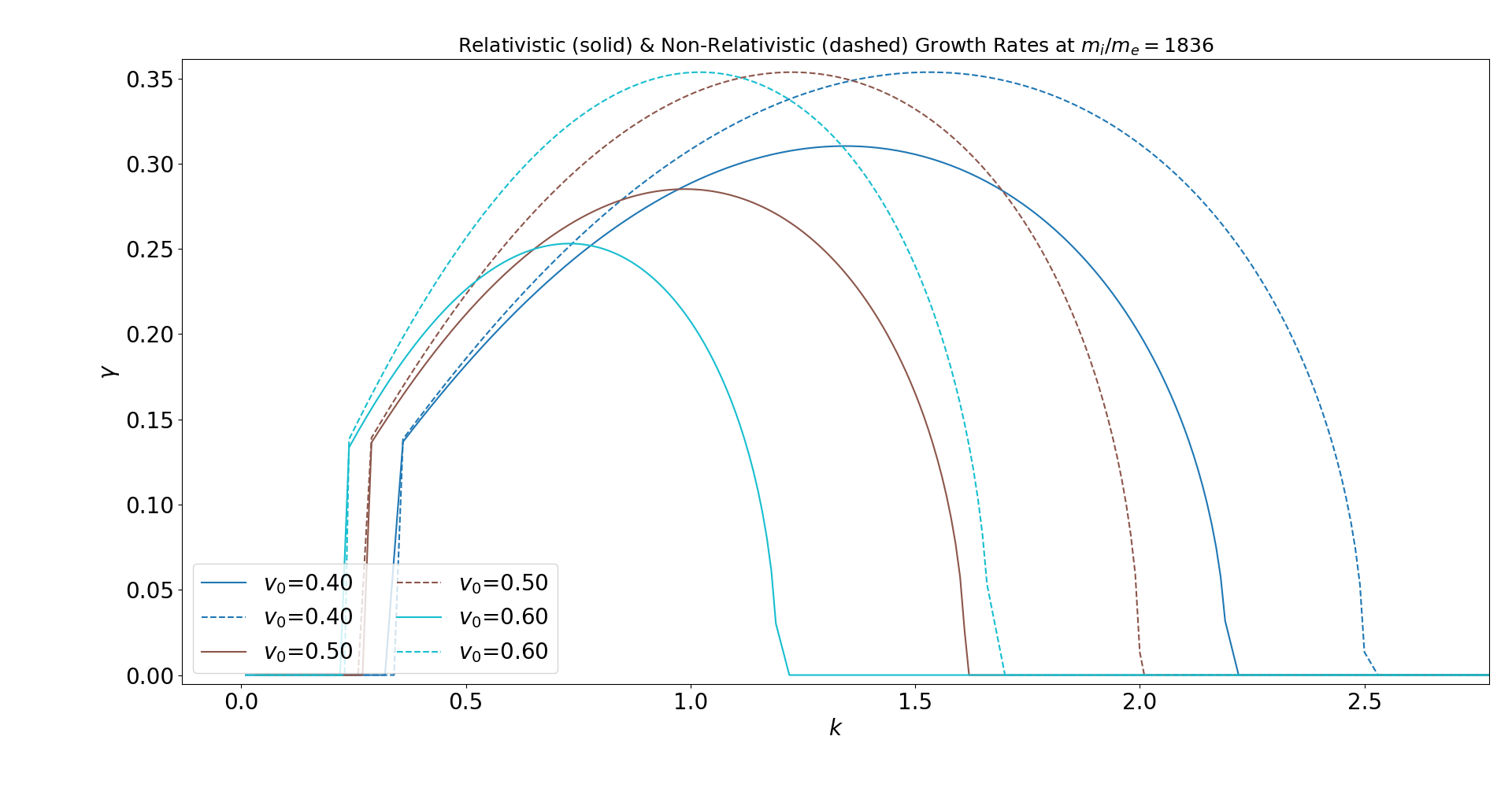}
    \end{subfigure}
     \hfill
    \begin{subfigure}[b]{0.4\textwidth}
        \includegraphics[width=\linewidth]{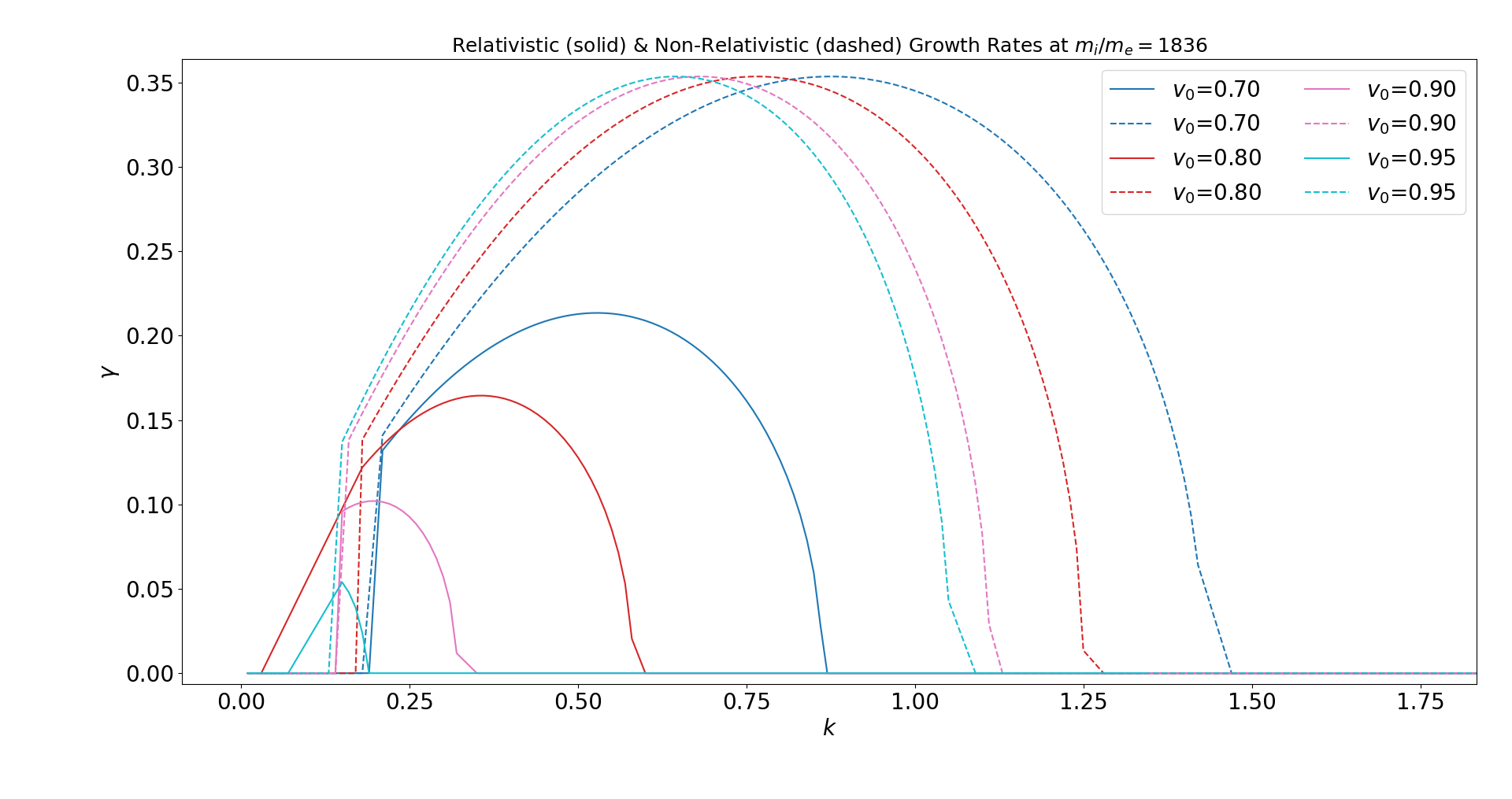}
    \end{subfigure}
 \caption{Growth rate $\gamma$ as a function of wavenumber $k$ for symmetric $m_i/m_e = 1836$. Relativistic (solid lines) and non-relativistic (dashed lines) results are shown for a range of drift velocities $v_0 \in [0.1, 1.0]$.}
    \label{gamma_mi_me_1836}
\end{figure}

By considering the dynamics of beam electrons and ions, plasma electrons and ions, and referring to this as a multi-species system, we present the growth rate $\gamma(k)$ as a function of wavenumber $k$ for mass ratios $m_i/m_e=1$ (electron-positron plasma) and $1836$ (electron-proton plasma) in figures~\ref{gamma_mi_me_1} and \ref{gamma_mi_me_1836}, respectively.\\

In the case of $m_i/m_e=1$ (figure~\ref{gamma_mi_me_1}), for low values of $v_0$ the wave growth can spread across a wide range of $k$. As $v_0$ increases, both the peak growth rate and the width of the unstable $k$ band decrease. Relativistic effects become more prominent at higher $v_0$, leading to a downward and leftward shift in the instability peak. This behavior is consistent with the theoretical predictions for electron-positron beam interactions examined in three-dimensional simulations~\cite{Silva2003}, where similar relativistic streaming instabilities have been observed to affect the growth characteristics at ultrarelativistic speeds. Recent experimental validation of these effects has been achieved by Warwick et al.~\cite{warwick2017experimental}, who directly observed current-driven instabilities in neutral electron-positron beams, confirming the modified instability characteristics in symmetric mass systems predicted by our relativistic analysis.\\

In the case of $m_i/m_e=1836$ (figure~\ref{gamma_mi_me_1836}), the overall growth rates are significantly lower and occur at lower wavenumbers. As $v_0$ increases, although the maximum growth decreases, the wavenumber of maximum growth decreases. These findings align with the particle-in-cell simulations of Dieckmann et al.~\cite{dieckmann2006particle}, who investigated ultrarelativistic two-stream instabilities with realistic mass ratio $m_p/m_e = 1836$ at relative Lorentz factor $\gamma(v_b) = 100$. Their simulations demonstrated that the most unstable wavenumber scales as \(k_u \approx \frac{\omega_{e1}}{v_b}\),
with the real part of the most unstable frequency $\omega_u \approx \omega_{e1}$, and growth rate 
\begin{equation}
\frac{\omega_i}{\omega_{e1}} \approx \frac{ \left( \frac{3\sqrt{3}\,\omega_{e2}^2}{16\,\omega_{e1}^2} \right)^{1/3} }{\gamma^{2/3}(v_b)},
\end{equation}
as reported in Dieckmann et al.~\cite{dieckmann2006particle}. The observed reduction in growth rates at higher streaming velocities was experimentally supported by their PIC simulations, which showed that electrostatic modes saturate through electron trapping mechanisms, with the growth and saturation occurring over similar timescales ($\omega_i^{-1}$) regardless of the beam density ratio, confirming the scaling behavior predicted by the relativistic dispersion relation~\cite{dieckmann2006particle}.\\

Energy transfer with respect to figures~(\ref{gamma_mi_me_1}) and (\ref{gamma_mi_me_1836}) reflects enhanced energy transfer from the electron beam to the plasma wave as the beam becomes more relativistic, although the ion response remains limited by its inertia. Overall, the instability exhibits strong dependence on both $v_0$ and the mass ratio. While higher drift velocities generally enhance the instability, the presence of a heavy ion species suppresses its magnitude and extent, especially at low $v_0$.\\

\begin{figure}[h!]
    \centering
    \includegraphics[width=0.5\textwidth]{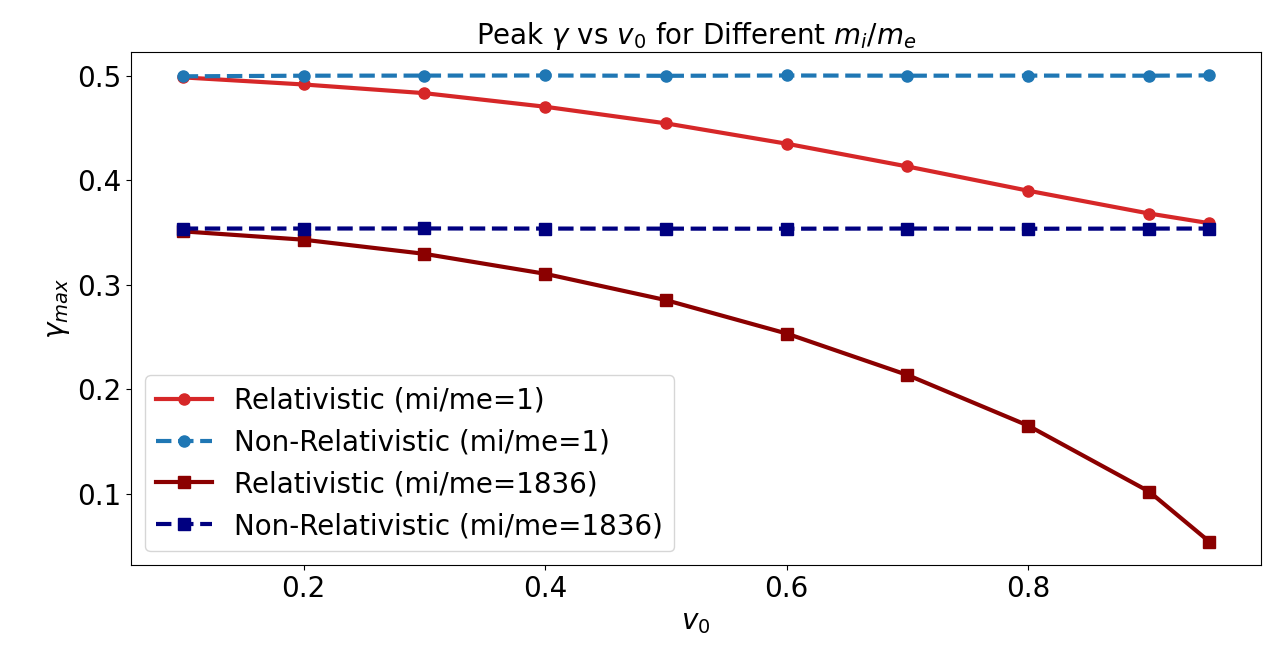}
    \caption{Maximum growth rate $\gamma_{\mathrm{max}}$ as a function of beam velocity $v_0$ for a two-species beam-plasma system with mass ratios $m_i/m_e = 1$ and $1836$.}
    \label{gamma max vs v0 multi}
\end{figure}

Figure~\ref{gamma max vs v0 multi} presents the maximum growth rate $\gamma_{\mathrm{max}}$ as a function of beam velocity $v_0$ for a two-species beam-plasma system, with ion-to-electron mass ratios $m_i/m_e = 1$ and $1836$. Maximum growth happens at $v_0=0.1$ due to enhanced beam-plasma resonance in both relativistic and non-relativistic cases. In the non-relativistic case, the maximum growth rate is independent of $v_0$ as evident from figure~\ref{gamma max vs v0 multi}. In the relativistic case, $\gamma_{\mathrm{max}}$ decreases with $v_0$ due to the increased effective inertia of the beam particles, captured by the Lorentz factor $\gamma_0(v_0)$. This decrease is sharper for $m_i/m_e=1836$. Therefore, the non-relativistic model fails to account for the suppression of growth rate, overestimating it at high velocities and resulting in an unphysical saturation as $v_0 \to 1$. The effect of the mass ratio is also evident: systems with smaller $m_i/m_e$ show stronger instabilities due to higher ion mobility, while for $m_i/m_e = 1836$, the ion effect is significantly reduced. These results highlight the importance of including relativistic corrections when modeling high-velocity beams, particularly in systems with lighter ions where the instability is more pronounced. The predicted relativistic suppression effects have been experimentally confirmed in laser-plasma acceleration studies~\cite{huang2017relativistic}, where instability suppression was observed in relativistic regimes, with preformed plasma channels providing restoring forces that validate our theoretical predictions of reduced growth rates at high velocities.\\

\begin{figure}[h!]
    \centering
    \includegraphics[width=0.5\textwidth]{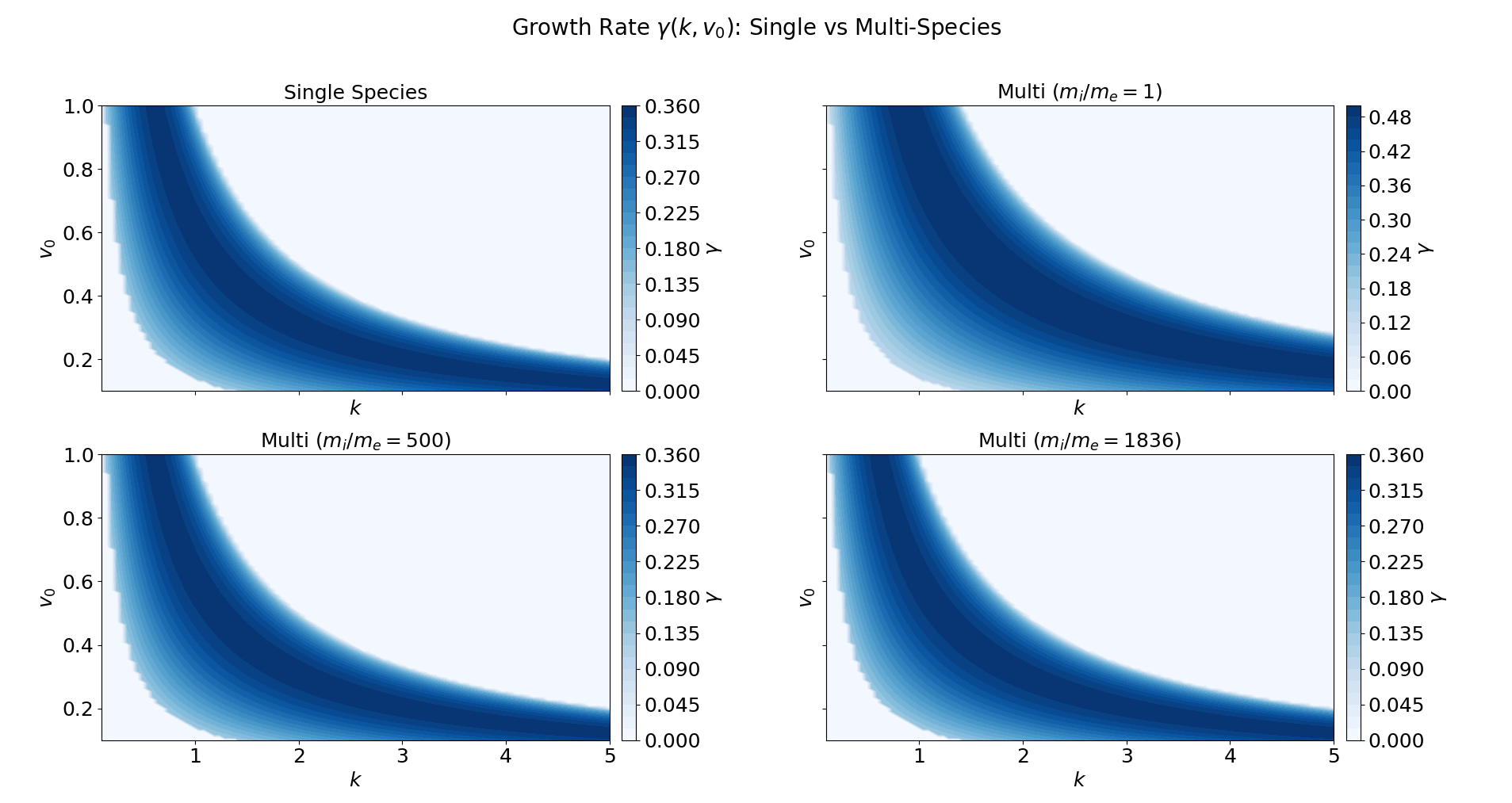}
    \caption{Growth rate $\gamma$ versus $(k, v_{0})$ for non-relativistic single species and multi-species ($m_i/m_e = 1$, $500$, and $1836$) plasmas.}
    \label{multi-com-non-rel}
\end{figure}

Figure~\ref{multi-com-non-rel} illustrates the contour plots of the growth rate  $\gamma$ versus $(k, v_{0})$ for non-relativistic single species and multi-species ($m_i/m_e = 1$, $500$, and $1836$) plasmas. In the single species case, the instability spans a broad region in $(k, v_0)$ space with maximum growth happening in the intermediate wavenumber range ($k \sim 1$--$3$). In lower and intermediate ranges, the growth rate increases with $v_0$. Almost the same pattern of growth rate happens in the multi-species case, except for $m_i/m_e=1$, where due to higher mobility of ions, the growth rate space spans more in both $k$ and $v_0$ dimensions.\\

\begin{figure}[h!]
    \centering
    \includegraphics[width=0.5\textwidth]{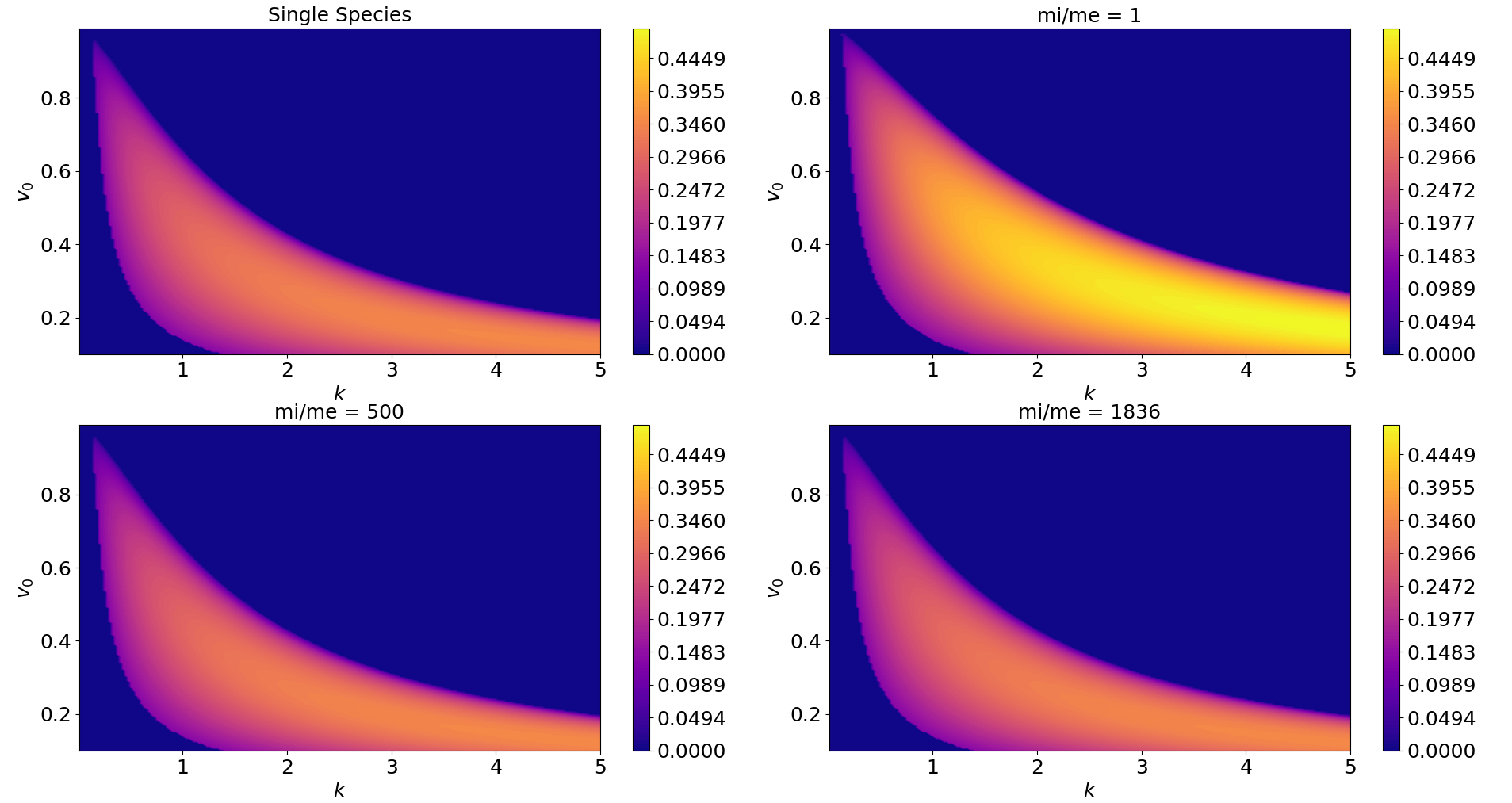}
    \caption{Growth rate $\gamma$ versus $(k, v_{0})$ for relativistic single species and multi-species ($m_i/m_e = 1$, $500$, and $1836$) plasmas.}
    \label{multi-com-rel}
\end{figure}

In the relativistic case (figure~\ref{multi-com-rel}), the span of growth rate in $(k,v_0)$ space follows the same pattern as in the non-relativistic case. However, the relativistic case shows that the instability onset is shifted to higher velocities due to the Lorentz factor $\gamma'_0$, which effectively reduces the contribution of high-speed beams to the dispersion relation. Although the maximum growth rate is generally lower due to the ${\gamma'_0}^{-3}$ scaling, the instability persists for higher $v_0$, and its peak shifts toward larger $k$ values. For all mass ratios, the unstable region in the relativistic case is narrower, and the influence of increasing $m_i/m_e$ is less pronounced compared to the non-relativistic case. Notably, for $m_i/m_e = 1$, the growth rate profiles resemble those of the single-species plasma, but for realistic mass ratios like $1836$, ion effects suppress the growth more effectively. These results highlight the necessity of relativistic corrections in analyzing high-energy beam-plasma interactions and demonstrate how species composition and mass asymmetry critically shape the instability landscape.\\

\section{\label{conclusion}Conclusion}

In this study, we have systematically investigated the influence of relativistic effects and multi-species plasma composition on the linear growth and structure of the two-stream instability. Beginning with the classical single-species model, we extended the dispersion relation to include fully relativistic corrections and contributions from ion species with varying mass ratios, including the realistic proton-electron case ($m_i/m_e = 1836$). Our numerical analysis revealed that relativistic effects significantly suppress the growth rate, particularly at high drift velocities, due to the ${\gamma'}_0^{-3}$ scaling of the plasma response. In the multi-species case, increasing ion mass further narrows the unstable region in $(k, v_0)$ space and reduces the maximum growth rate. This stabilizing influence is especially pronounced in the relativistic regime, where both electron and ion inertia affect the resonance condition and the overall instability structure. Contour plots and growth-rate profiles demonstrated how both physical ingredients- relativity and ion dynamics- alter the threshold and efficiency of instability. The parameter space analysis revealed that the instability is increasingly confined to lower wavenumbers and narrower velocity ranges as relativistic effects become dominant, with multi-species configurations providing additional stabilization through enhanced ion inertia contributions. Our comparative study across all four regimes shows that neglecting either relativistic corrections or realistic mass ratios leads to significant overestimation of growth rates and unstable spectral ranges, particularly in high-energy applications. The results underscore the limitations of simplified models and highlight the necessity of incorporating both relativistic and multi-species effects in the accurate modeling of high-energy plasmas. Recent experimental advances provide strong validation for our theoretical predictions. Laboratory creation of relativistic electron-positron plasmas~\cite{warwick2017experimental} indicates similar results as that of our symmetric mass ratio results, while laser-plasma acceleration experiments~\cite{huang2017relativistic} have demonstrated the instability suppression effects indicated by our relativistic analysis. These experimental developments, combined with observations of stable astrophysical jets that classical theory cannot fully explain, strongly support our findings that both relativistic corrections and realistic mass ratios are essential for accurate modeling of two-stream instabilities in high-energy plasma systems.


\begin{acknowledgments}
Authors VS and MKC are grateful to the
 University Grants Commission, India for providing a Non-NET fellowship. VS and RM thanks James Juno and Ammar Hakim at Princeton Plasma Physics Laboratory and Petr Cagas at Virginia Tech for helpful suggestions and discussions. Visiting associateship program of Inter-University Centre for Astronomy and Astrophysics (IUCAA) is acknowledged by RM. All simulations have been performed in `Brahmagupta' HPC facility at Sikkim University. 
\end{acknowledgments}

\section*{\label{references}References}
\bibliography{aipsamp}

\end{document}